\documentclass[iop,revtex4]{emulateapj}
\usepackage{amssymb,stackengine}
\usepackage{xcolor}
\definecolor{rkka}{RGB}{219,66,32}

\begin{document}

\newcommand{\kms}{km~s$^{-1}$\,}
\newcommand{\msun}{$M_\odot$\,}
\newcommand{\eb}{\begin{equation}}
\newcommand{\ee}{\end{equation}}

\renewcommand{\topfraction}{1.0}
\renewcommand{\bottomfraction}{1.0}
\renewcommand{\textfraction}{0.0}

\title{Gyr-timescale destruction of high-eccentricity asteroids by spin and why 2006 HY51 has been spared}
\shorttitle{Spin-up of high-eccentricity asteroids}

\author{Valeri V. Makarov}
\affil{U.S. Naval Observatory, 3450 Massachusetts Ave., Washington, DC 20392-5420, USA}
\email{valeri.makarov@gmail.com}

\author{Alexey Goldin}
\affil{Teza Technology, 150 N Michigan Ave, Chicago IL 60601, USA}
\email{alexey.goldin@gmail.com}

\author{Dimitri Veras}
\affil{Centre for Exoplanets and Habitability, University of Warwick, Coventry CV4 7AL, UK\\
Department of Physics, University of Warwick, Coventry CV4 7AL, UK}
\email{d.veras@warwick.ac.uk}
\thanks{STFC Ernest Rutherford Fellow}

\begin{abstract}
Asteroids and other small celestial bodies have markedly prolate shapes, and the
perturbative triaxial torques which are applied during pericenter passages in highly eccentric
orbits trigger and sustain a state of chaotic rotation. Because the prograde spin rate around the
principal axis of inertia is not bounded from above, it can accidentally reach the
threshold value corresponding to rotational break-up. Previous investigations of this
process were limited to integrations of $\sim 10^3$ orbits because of the stiff equation of motion.    
We present here a fast 1D simulation method to compute the evolution of this spin rate
over $\sim 10^9$ orbits. We apply the method to the most 
eccentric solar system asteroid known, 2006 HY51 (with $e = 0.9684$), and find
that for any reasonably expected shape parameters, it can never be accelerated to break-up speed.
However, primordial solar system asteroids on more eccentric orbits may have already broken up
from this type of rotational fission. The method also represents a promising opportunity to 
investigate the long-term evolution of extremely eccentric triaxial exo-asteroids ($e > 0.99$), 
which are thought to be common in white dwarf planetary systems.
\end{abstract}



\section{Introduction}
\label{sec:intro}

Minor planets on highly eccentric orbits ($e \gtrsim 0.9$) play important roles during
the formation and death of planetary systems. These objects
contributed to the lunar bombardment \citep{morbetal2018,braetal2020} 
and to the accretion of primordial terrestrial planets, both within the solar
system \citep{obretal2018} and in extrasolar planetary systems such as
TRAPPIST-1 \citep{denreg2019}. Around white dwarf planetary systems, highly 
eccentric asteroids are thought to be the primary progenitor of debris disks 
\citep{jura2003,debetal2012,veretal2014,malper2020a,malper2020b}
and observed metallic pollution in the photospheres of those stars 
\citep{zucetal2010,koeetal2014}.

Although often represented as point masses in numerical simulations, most known
asteroids are aspherical. This asphericity significantly affects their spin dynamics, which
is coupled with their orbital evolution. 

The shape and distribution of mass of a rocky asteroid can be idealized
and approximated with an ellipsoid with three unequal principal axes. In dynamical
problems, the most important parameters of this approximation are the
three corresponding moments of inertia $A$, $B$, and $C$, in increasing order.
They can be related to the geometric elongation parameters under the assumption
of uniform density, or by direct integration for a given density profile
and specific multi-layer models. 

It is a well known fact that the size and mass
of celestial bodies is strongly correlated with the degree of triaxiality
(and, generally, with deviations from spherical symmetry), which can be
represented by the dimensionless ratio $\sigma=(B-A)/C$. The benchmark value
is $\sigma=1.9\times 10^{-5}$ for Earth \citep{lam80}, with its mass $M_{\oplus}=5.97\times
10^{24}$ kg. Larger super-Earth exoplanets are likely to have shapes that are
even closer to perfect spherical symmetry, while the Moon, with a mass that is
roughly 100 times smaller ($7.35 \times 10^{22}$ kg) than Earth's, has a more prolate shape
at $\sigma=2.28 \times 10^{-4}$. Other satellites of similar mass  have even
greater degrees of deformation, e.g., Io with $\sigma=6.4 \times 10^{-3}$
\citep{and}. 

This inverse correlation between mass and $\sigma$ continues into the domain 
of comets and asteroids,
where irregular potato-like shapes are prevalent. The recently discovered
asteroid 'Oumuamua \citep{meeetal2017,williams2017}, which is probably of interstellar origin \citep{hig},
represents an extreme case with an observable aspect ratio of $6\pm1$ \citep{mcn}, which is
likely to be even greater because of projection effects \citep{sir}. 

When an elongated body in an eccentric orbit passes close to its host\footnote{If the elongated
body is an asteroid or comet, the host is the parent star. The elongated body may also be a satellite,
in which case the host is the parent planet.}, the former is subject to the triaxial
torque of increasing magnitude due to an increasing gradient of the gravitational potential.
The resulting equations of motion (Euler's equations) have been well-established and
represent three second order nonlinear differential equations, which include
the instantaneous direction cosines of the longest axis and the instantaneous spin
rates about all three principal axes \citep{dan}. The system simplifies into a single second
order ordinary differential equation (ODE) under the assumption that the principal axis 
of inertia (corresponding to the moment $C$) is always orthogonal to the orbital plane, 
which nullifies the other two components of the torque and the velocity-dependent terms. 

In this paper, we provide a computational method which enables long-term tracking
of the spin evolution of elongated bodies on highly eccentric orbits. The 1D spin evolution
model is equivalent to that in \cite{mav}. In that paper, however, evolution was computed
on a timescale which is roughly six orders of magnitude shorter than the current age of the
solar system. Our method enables us to support the primordial dynamical status of
2006 HY51, the asteroid whose orbit is of the highest known eccentricity.

In Section 2, we summarize our knowledge of the most eccentric solar system asteroids
and provide more details about the spin evolution.
We then describe our numerical integration method in Section 3, before identifying
the origin of chaos in our systems in Section 4 and integrating the evolution of 2006 HY51
in Section 5. We discuss and summarize our results in Section 6.

\section{The highest eccentricity asteroids in the Solar system}
\label{ast.sec}
The JPL Horizons database provides a compendium of information about celestial
bodies in the Solar system, which is available online\footnote{\url{https://ssd.jpl.nasa.gov/?horizons}}.
We made a selection of all Sun-orbiting asteroids in this database with
semimajor axes $a < 3$ au, absolute magnitudes $H_{\rm mag} < 20$, and eccentricities
$ 0.95 \le e < 1.0$ (filtering out objects on hyperbolic orbits with eccentricity
above 1). 

The selected 6 objects are listed in Table~\ref{obj.tab}. They do not
appear to have attracted much attention in the literature, except as representing 
potentially hazardous Near-Earth Objects. Although these objects are 
usually faint with visual magnitudes in the 23--25 range, they become bright during
brief perihelion flybys. The most eccentric orbit is found for 2006 HY51, which is 
arguably the most studied object on the list. It is the
only one with an estimated diameter, which is 1.218 km. Its rate of rotation is unknown.

\begin{deluxetable*}{r r r r l } 
\tablecaption{High-eccentricity asteroids. \label{obj.tab}}
\tablewidth{0pt}                                   
\tablehead{    
\colhead{Name} & 
\colhead{$e$} & 
\colhead{$a$} & 
\colhead{$r_{\rm min}$} &
\colhead{$P_{\rm orb}$} \\
\colhead{} & \colhead{} & \colhead{au} & \colhead{au} & \colhead{yr}
}
\startdata
394130 (2006 HY51)  & 0.9684 & 2.590 & 0.082 & 4.17 \\
(2011 KE) & 0.9546 & 2.206 & 0.100  & 3.28 \\
465402 (2008 HW1)  & 0.9600 & 2.586 & 0.103 &   4.16 \\
(2012 US68) & 0.9579 & 2.504 & 0.105  & 3.96 \\
399457 (2002 PD43)  & 0.9560 & 2.508 & 0.110 & 3.97 \\
431760 (2008 HE)  & 0.9505 & 2.261  & 0.112  & 3.40 \\
\enddata
\end{deluxetable*}

These six asteroids of the Apollo group have approached the Sun within 0.1 au 
every 3 or 4 years over most of the solar system's lifetime (billions of years). 
The spin of these objects, driven by the periodic flybys, is chaotic.
Chaotic rotation of prolate asteroids is predicted theoretically \citep{wis84},
as well as observed in the solar system \citep{kou}. Even if some of the
currently synchronized planetary satellites of elongated shape, such as Phobos and
Epimetheus, stay within an island of stable libration, they could not have reached
this state without crossing a zone of chaotic motion in the past \citep{wis87}.

Minor bodies in highly eccentric orbits, on the other hand, are permanently
in the chaotic state of rotation driven by the impulsive kicks during brief
periapse phases \citep{mav}. The rate of rotation around the principal axis
of inertia is a continuous function of time with step-like changes, which can be
considered as a random process when discretized at
apoastron times. 

At a given apoastron rotation velocity, as demonstrated with
direct numerical integrations, the distribution of velocity updates is a concave function
of the orientation angle. This distribution is also biased with respect to zero in such a way
that the greatest positive updates are larger in absolute value than the greatest negative updates
at small and moderate apoastron spin rates, and the opposite bias is present for high prograde
spin rates. This curious property makes the chaotic evolution weakly stationary and self-regulating,
so that if the spin rate stochastically becomes very high, the updates are more likely to
reduce it. Even though the velocity is not bound on the high end, the regulation mechanism
may require a long time for the process to reach the break-up spin, depending mostly on
orbital eccentricity. 

This property of the long-term evolution of the regulation mechanism is the primary
motivation for this work. Although rotational break-up over just $10^3$ orbits has been 
invoked to generate the ring of debris material orbiting white dwarf ZTF J0139+5245 
\citep{vanetal2019,ver+20}, the short timescale chosen for the integrations restricted
the parameter space which could be explored. Integrations which last over
the lifetime of the solar system now allows us to probe the evolution of solar system asteroids
like 2006 HY51 and help assess if they could be primordial.

\section{Numerical integration of motion}

We adopt the 1D model described in \cite{mav}, which uses a simplified equation of
motion with the axis of rotation aligned with the principal inertia axis ${\mathcal C}$ and
orthogonal to the orbital plane at all times. The torque acting on the asteroid is then
parallel to ${\mathcal C}$. This assumption is realistic when the asteroid is spinning fast in the
prograde sense so that its angular momentum is aligned with the orbital angular momentum,
and free librations around the other axes of inertia can be ignored. We performed
limited integrations (which are much more computer-intensive) of the full triaxial equation
of motion with random initial conditions. These simulations confirmed that the principal axis
rotation remains strongly chaotic and driven by impulse-like interactions at perihelia. We surmise
that the time scale of spin evolution may become somewhat longer due to the geometric projection
of the longest axis at nonzero inclination, but this needs to be confirmed by extensive
computer modelling.

The sidereal orientation angle $\theta$ in the equatorial plane of the asteroid is
convenient to measure from the direction to the perturber's periapse (perihelion in our case)
and the longest axis of the ellipsoid ${\mathcal A}$. 
A second-order ODE includes $\theta$, its first time derivative $\dot\theta = \omega$, and its
second time derivative (rotational acceleration), as well as the triaxiality parameter
$(B-A)/C\equiv \sigma$, the orbital mean motion $n$, the semimajor axis $a$,
the instantaneous orbital distance $r$,
and the true anomaly $f$. Two initial conditions are arbitrarily chosen at times $t=0$ for both
$\theta$ and $\omega$. These conditions are sufficient to integrate this equation in time 
and to determine the functions $\theta(t)$ and $\omega(t)$. We further compute the true anomaly 
from the mean anomaly and the eccentric anomaly, which is also needed to compute $r$, 
via Kepler's equation and reverse interpolation. 

This integration requires special care for
high values of eccentricity because of rapid, large amplitude variations of the integrated parameters
at short perihelion passages, where the polar torque suddenly increases by several orders of
magnitude. This problem is therefore stiff, requiring that the integration algorithm adopt 
an un-economical very small integration time step, or adjust the integration step
according to the local stiffness.

\begin{deluxetable*}{l c c l } 
\tablecaption{Observed and assumed parameters of 2006 HY51. \label{param.tab}}
\tablewidth{0pt}                                   
\tablehead{    
\colhead{Parameter} & 
\colhead{Value} & 
\colhead{Unit} & 
\colhead{Origin}
}
\startdata
Period, $P_{\rm orb}$   & 4.17 & yr & observed \\
Mean motion, $n$  & 0.00413 & rad d$^{-1}$  & observed \\
Semimajor axis, $a$  & 2.59  & au  & observed \\
Eccentricity, $e$  & 0.9684  &   & observed \\
Mass, $M_2$  & $7.6\times 10^9$ & kg  & assumed \\
Radius, $R$  & $0.609\times 10^3$ & m  & observed \\
Triaxiality, $\sigma=(B-A)/C$ & 0.2 & & assumed \\  
\enddata
\end{deluxetable*}

\begin{figure}
\plotone{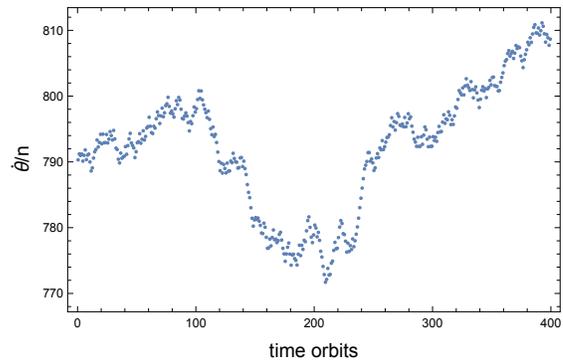}
\caption{Simulation of spin rate in units of mean motion of the 2006 HY51 asteroid for 400 consecutive orbits
by small time-step integration with stiffness switching. 
Only the aphelion values of rotation rate are shown with dots. 
\label{cha.fig} }
\end{figure}

Integration of the ODE with the parameter values for 2006 HY51 (listed in 
Table~\ref{param.tab}) confirms that the principal axis rotation of this object is 
strongly chaotic because of
abrupt changes of $\omega$ at perihelia of seemingly random magnitudes and direction.
If we only record the values of $\omega=\dot\theta$ at the times of aphelia, then the stochastic
changes of this parameter -- an example of which is shown in Fig. \ref{cha.fig} -- can be
formally fitted as a discrete random process, similar to a random walk. 

These fits using general random process models are not very successful because 
they do not capture peculiar properties of the stochastic behaviour \citep{mav}. Specifically, the
velocity differences or ``updates" between consecutive aphelion passages, $d_{\omega, i} =
\omega_i - \omega_{i-1}$, for a fixed $\omega_{i-1}$, do not follow a Gaussian distribution,
or even a bell-shaped distribution. The spin updates $d_{\omega, i}$ in  Fig. \ref{cha.fig} 
are relatively small, resulting in a fairly slow variation of the spin rate, because of the 
strong statistical dependence of this parameter on the spin rate itself. 

\begin{figure}
\plotone{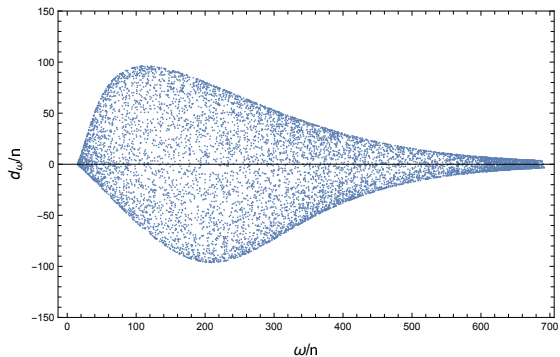}
\caption{Spin rate updates versus spin rates at aphelion for $50\,000$ numerically integrated orbits of 2006 HY51
with $\sigma=0.2$. 
\label{bub.fig} }
\end{figure}

We performed a series of exact long-term integrations (which are much slower than our new
method) of 9000 orbits each for 2006 HY51 with
different initial parameters and recorded the tuples $\{\omega_{i-1}, d_{\omega, i}\}$
for each orbit. The result of one such exact integration is shown in Fig. \ref{bub.fig}.
Each dot represents one aphelion tuple. The occupied area in the parameter space is similar
in shape to the distributions found for exo-asteroids orbiting white dwarfs at higher eccentricity
\citep{mav}. The range of possible updates at a fixed aphelion velocity is finite with sharp
boundaries, which are not symmetric around zero. 
The upper boundary of positive updates is
markedly greater in absolute value than the lower boundary for negative updates at $\omega
\lesssim 100\;n$. Therefore, the spin rate will not stay in this low-value regime for a long time
and inevitably will meander to higher values. The opposite asymmetry is found at $\omega
\gtrsim 200\;n$, and although $\omega$ is not bounded on the high end, the general trend
is to stochastically slow down. 

This peculiar probability distribution of velocity updates makes the process somewhat self-regulated.
In the context of this study, we are interested in the high spin regime of the process. What is the likelihood
of the spin rate achieving very high prograde values? 

The narrow funnel at the high end
of the distribution stretches to infinity (Fig.~\ref{bub.fig}), but the range of possible updates also infinitely
decreases, maintaining a tiny negative bias. Therefore, when the spin rate becomes higher than
several hundred $n$, the updates will be small and the asteroid will become stuck in this area 
for an extended interval of time, with the probability of spinning further up becoming 
negligibly small. Indeed,
this specific realization failed to reach spin rates much above $700\,n$
in $9\,000$ orbits. Given a much longer time, it is a matter of probabilities that the spin rate increases
to, say, $1000\,n$. We hence would like to answer these questions: how likely is it for 2006 HY51 to reach
the fission threshold within a certain time span; or, alternatively, what is the characteristic time
of reaching the fission velocity at a certain probability?

Although these questions can be answered through expensive numerical simulations, such 
direct in-orbit integrations are too computer-intensive and slow. The limited integrations 
presented in Figs. \ref{cha.fig} and \ref{bub.fig}
indicate that it may be impossible for 2006 HY51 to ever reach the fission velocity within the solar system
age. To confirm and verify this surmise, we look for a surrogate simulation model representing this stochastic
process. The idea is to compute the velocity update $d_{\omega, i}$ directly from tuples $\{\theta_{i-1},
\omega_{i-1}\}$ for each orbit avoiding the in-orbit integration of the equation of motion.

\section{The origin of chaos}
\label{cha.sec}

We consider the function $d_\omega(\theta,\omega)$ (herewith we drop the indices for brevity) and its
properties. This function is deterministic, i.e., given a pair of specific aphelion parameters, it takes
one exact value, which can be accurately computed. This feature may seem to contradict the 
obviously chaotic character of the process, but the model function does not have to be stochastic 
for chaos to emerge. For example, the function may have discontinuities or singularities, but not 
for asteroid rotation (at least, when the parameters are not too close to a separatrix). 

\begin{figure}
\plotone{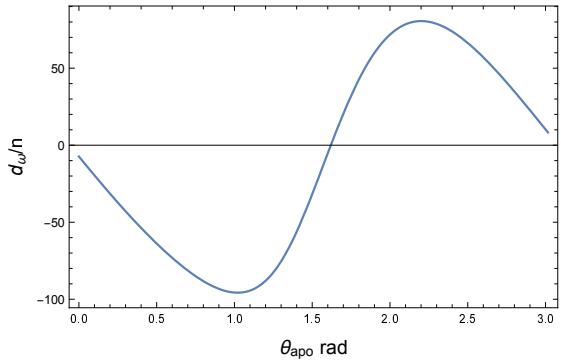}
\caption{Spin rate update $d_\omega/n$ as a function of aphelion orientation angle $\theta_{\rm apo}$
at a fixed aphelion spin rate $\omega_{\rm apo}/n=200$.
\label{dom.fig} }
\end{figure}

Fig. \ref{dom.fig}
shows the function values computed by full-scale integration for a fixed aphelion velocity $\omega
=200\,n$ and a grid of orientation angles $\theta$ covering the entire range $[0,\pi]$. The function is
a remarkably smooth and simple periodic function with two dominating harmonics $\sin 2\theta$ and
$\sin 4\theta$, and a constant term. Interpretation of this function is intuitively clear: at a fixed
aphelion velocity, the object rotates almost uniformly on the way to perihelion because the torque
is numerically small at a great distance from the Sun until it reaches the point of perihelion, where
the violent interaction takes place within a very short interval of time. The object will 
make a certain number of revolutions between the aphelion and perihelion, which is weakly 
dependent on the initial orientation angle.
But the outcome of the perihelion interaction is strongly (almost exclusively) dependent on the orientation
angle at perihelion. By changing the aphelion $\theta$ we change the perihelion $\theta$ by nearly the same
amount, and variation in $\theta$ within $1\,\pi$ samples the entire range of possible velocity
updates. 

In another experiment, we fix the aphelion angle at a certain value and change the aphelion velocity $\omega$
in the vicinity of the initial $\omega_0=200\,n$.
Numerical integrations confirm that {\it nearly} the same range of  $d_\omega$ is sampled when 
$\omega$ varies within
$[-1/2\,n,+1/2\,n]+\omega_0$. Furthermore, the curve $d_\omega(\omega)$ looks remarkably similar to the dependence
$d_\omega(\theta)$ on these intervals, apart from an additional slow trend in the former. The reason for this 
similarity is that the two effects are interchangeable, i.e., the same value of perihelion orientation 
can be achieved by either adjusting the aphelion orientation within $\pi$ or by adjusting the aphelion 
velocity within $1\,n$. 

We can accurately map the 2D function $d_\omega(\theta,\omega)$ on a grid of
points covering the domain of interest, which is between the minimum ``stop" velocity 
($\simeq 20\,n$, according to
Fig. \ref{bub.fig}) and the fission critical velocity, and 0 to $\pi$ in angle. This function takes high-frequency
wiggles in the $\omega$ dimension, but is smooth and sinusoid-like along $\theta$ almost everywhere, except
in the corner close to the minimum velocity. We speculate that in that area of the parameter space, the
system has to cross the separatrix to probe very slow prograde or retrograde velocities.

Where does the explicitly chaotic behaviour come from in this system with a smooth and integrable model function?
The cause of chaos in this case is the large sensitivity of the aphelion angle to the preceding perihelion
velocity update. The asteroid makes multiple rotations on each half-orbit, to the effect that even a small
change in $\omega$ can sample the entire range of $\theta$. Therefore, the velocity update at each perihelion
becomes almost independent of the previous perihelion configuration and the previous velocity update.
This explains why the sequence $d_{\omega, i}$ is best approximated with a GARCH$(1,1)$ random process
with the current variance strongly dependent on the preceding variance but weakly dependent on the
preceding value \citep{mav}. There is no obvious correlation with the state just two orbits ago, because
the aphelion orientation angle is effectively scrambled at each periastron passage.

\section{Long-term simulations of 2006 HY51}
\label{long.sec}

The smoothness of the model function $d_\omega(\theta,\omega)$ opens up the possibility to replace the
cumbersome small-step integration of the equation of motion with a very fast and efficient simulation
of tuples $\{\omega_{i-1}, d_{\omega, i}\}$. We replace this function with a 2D interpolation function
defined on a grid of points in a rectangular area in the $\{\theta,\omega\}$ plane. 

\begin{figure}[h]
\plotone{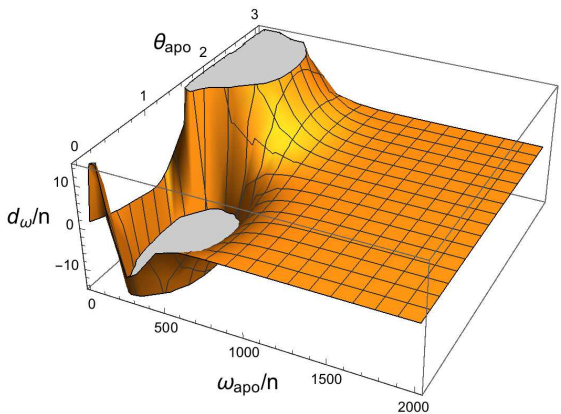}
\caption{Interpolation function of spin rate updates $d_\omega(\omega_{\rm apo}, \theta_{\rm apo})/n$
estimated from grid integration. The extrema of the function are clipped to better show the complex
behavior at the lower boundary of spin rate, $\omega_{\rm apo}/n\simeq20$.
\label{int.fig} }
\end{figure}

Fig. \ref{int.fig}
shows such an interpolation function computed on a grid of nodes separated by $\pi/10$ in $\theta$
and $20\,n$ in $\omega$. This function is very smooth everywhere except in the close vicinity of the
lower bound of velocity. It does not, however, capture the waves  between the interpolation nodes
in the $\omega$ dimension, which have a period of $1\,n$. We will explain now why this smoothed
version of the model function is sufficient for accurate simulation of the rotation process.

As we discussed in Sect. \ref{cha.sec}, the process of rotation at high eccentricity is almost memory-less,
in that each aphelion orientation is barely correlated with the previous aphelion state. Therefore,
without a loss of fidelity, the aphelion orientation $\theta$ can be randomized, i.e., drawn from
a uniform distribution over $[0,\pi]$ at each orbit. The simulation process starts with some initial
values $\{\theta_1,\omega_1\}$ and the corresponding $d_\omega$ is computed from the interpolation
function shown in Fig. \ref{int.fig}. The next state is computed simply as $\omega_2=\omega_1+d_\omega$,
while $\theta_2$ is a random number between 0 and $\pi$. This randomization of $\theta$ makes it
unnecessary to take into account for the 20 waves of the actual model function between the nodes,
because the computed $d_\omega$ values are drawn from the same distribution while the interpolation
function accurately captures the smooth part of the velocity-dependent variation. 

\begin{figure}
\plotone{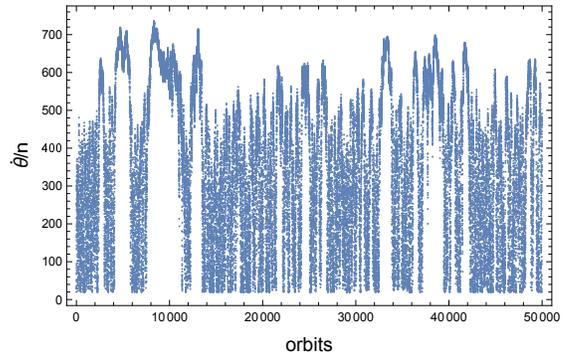}
\caption{Chaotic evolution of the 2008 HY51 asteroid rotation rate simulated using our fast
Monte-Carlo method. Only a small section of $50\,000$ orbits from a much longer
integration span is shown.
\label{sim.fig} }
\end{figure}

By using this fast simulation method, we performed 100 Monte-Carlo simulations of rotation of the
2006 HY51 for $5\times 10^6$ orbits each, with the parameters listed in Table \ref{param.tab}. This
ensemble is practically equivalent to a single simulation of $5\times 10^8$ orbits, or $2.085$ Gyr, 
because the initial
conditions for each trial can be arbitrarily chosen. One of the simulated rotation curves is shown in
Fig. \ref{sim.fig}. The spin rate rarely increases beyond $700\,n$, where the range of possible
updates shrinks almost to zero. The highest spin rate at aphelion achieved in this simulation is
$994\,n$, which corresponds to a period of 1.5 d. The median spin rate over half a billion orbits is
$369\,n$ ($P_{\rm rot}=4.1$ d). 

The unknown parameter $\sigma$, which defines the prolate shape of the asteroid, has the highest
significance for rotation evolution. In the previous set of trials, $\sigma = 0.2$ has been
used. To estimate the 
potential uncertainty in $\sigma$, we performed a similar
set of trials with $\sigma=0.4$. Exact integration for a much smaller duration showed that approximately
twice as large
velocity updates are possible, as expected, and that the asymmetry of the distribution presented
in Fig. \ref{bub.fig} is much more pronounced. More importantly for this study, the high-velocity
funnel is wider than in the $\sigma=0.2$ case. We should expect higher spin rates being achieved more easily.
However, the long-term simulations with the fast method reveal that the gain is rather modest. The median
velocity is $419\,n$ ($P_{\rm rot}=3.6$ d), and the absolute maximum is $1035\,n$ ($P_{\rm rot}=1.5$ d).

\section{Spin-orbit resonances}
To validate the results obtained with our fast tuple simulation method with random aphelion orientations,
we performed multiple, long-term simulations with a high fidelity algorithm which requires massive computations
on parallel computer cores. This modification\footnote{Our simulation code in Julia for this modification of rotation
simulation method is available online at \url{https://github.com/agoldin/2006HY51}} is based on a 
re-parameterization of the problem introducing
an auxiliary parameter $\theta^*=\theta_{\rm ap}+ \pi\,\omega_{\rm ap}/n$, which is numerically close to, but
better behaved, than the actual value of $\theta$ at perihelion. An interpolation function $d_\omega(\theta^*,\omega)$
is computed on a fine grid of nodes in both arguments. The main advantage of
this re-parameterization is a much smoother
interpolation function $\theta(\omega, \theta^*)$ without the high-frequency waves in the $\omega$ dimension. Using
the new $\theta_{\rm ap}(\omega, \theta^*)$ function in addition to $d_\omega(\theta^*,\omega)$ allows us to compute both
the velocity update and the next aphelion orientation angle for each orbit without replacing $\theta_{\rm ap}$ with
a randomly generated number. The downside of this method, besides its higher computing requirements,
is that the interpolation functions become shredded at the low-$\omega$ boundary, resulting in occasional
``diffusion" into the range of retrograde spins -- a process we never saw in our limited full-scale integrations.

With this high-fidelity algorithm, we performed 512 simulations of the spin rate process each covering
250 million orbits. The overall behaviour and statistics of these simulations are in excellent
agreement with the fast Monte-Carlo method, with one important difference. Only a small fraction of these trials
demonstrated chaotic walk through the end of the time span. The rest ended up in certain quantum states
in the parameter space characterized by nearly constant aphelion orientations and spin rates. These are
spin-orbit resonances similar to the ``islands" of stable rotation found by \citet{wis84}, but pushed
to the domain of very high velocities. We can estimate the characteristic time of capture into resonance
by counting the quantiles of the chaotic phase durations. Of the entire set of 512 simulations with random initial
conditions, 25\% were captured within $2.5\cdot 10^6$ orbits, and 50\% -- within $6.5\cdot 10^6$ orbits.
Thus, a significant fraction of trials show resonance capture within 20 to 50 million years. 
The resonance endpoints in $\omega_{\rm ap}/n$ are quantized with the lowest frequency around 960. We note that the resonance spin rates are only slightly lower than the
maximum values seen at the chaotic stages. The aphelion orientation angles are integer multiples of
$\pi/2$. 

\begin{figure}
\plotone{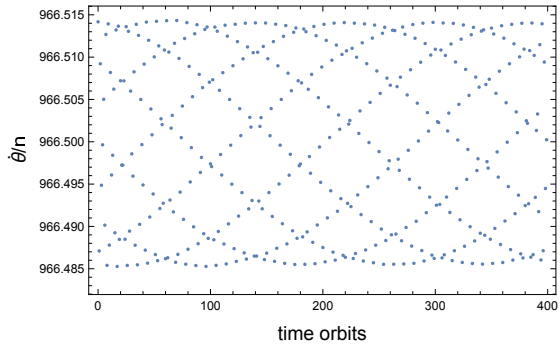}
\caption{Resonance rotation of the 2008 HY51 asteroid simulated with full-scale integration of
the equation of motion. Only aphelion values of spin rate are shown. The initial parameters
were found using long-term simulations with the high-fidelity interpolation technique (see text).
\label{res.fig} }
\end{figure}

We verified these results by taking one of the endpoints with $\theta_{\rm ap}=4.71115727$ rad, $\omega_{\rm ap}=
966.514171\;n$ (with $n=0.00412528$ rad d$^{-1}$, $\sigma=0.2$, $e = 0.9684$) and used it as initial condition
in numerical integration of 400 orbits. The output is depicted in Fig. \ref{res.fig}. The trajectory is periodic
circulation around a point of stable equilibrium with $\omega$ varying within a narrow range. Only one force is
in action in the system, and it may seem puzzling how this resonance can be stable in the absence of a restoring
force. We found a semi-qualitative interpretation, which also explains why the resonances are limited to the
range of extreme rotation velocities.   

The function $d_\omega(\theta,\omega)$ considered in \S\ref{long.sec} has multiple roots. Replacing
aphelion $\theta$ and $\omega$ with $x$ and $y$ for convenience, the differential updates to these
parameters are $\{dx,dy\}=f(x,y)$. The roots of $f$ form lines in the $\{x,y\}$ plane, which are spaced
by roughly $1\,n$ in the frequency dimension. When during the chaotic walk in the parameter space
the process hits one of the roots $\{x_0,y_0\}$, the updates are nullified and the process becomes
constant. For this state to be stable, a small departure from the exact root value $\delta$ should result in
an opposite-signed update, which should be smaller in absolute value than $2\,\delta$. We know that at high
prograde rates the function is approximately sinusoidal so we can write at fixed $x=x_0$ that $dy'(x_0,y_0+\delta)
=\Omega \sin(2 \pi \delta /n)$. Using the first-order Taylor expansion in $\delta$, $dy'=2\,\pi\Omega\,
\delta /n$ with a plus or minus sign for the two roots in $x$. Only the root with a minus sign is of interest, and the condition of stability is then
\eb
\Omega < n/\pi
\ee
We know from our numerical integrations that the amplitude of possible velocity updates becomes that small
only at high prograde rates of rotation (cf. Fig. \ref{bub.fig}). The limiting aphelion rate is approximately
$960\,n$. Multiple roots exist at lower rates but these equilibria are inherently unstable because the updates
are greater than the perturbation.

\section{Discussion and conclusions}
\label{dis.sec}

We have presented a fast method to compute the long-term spin evolution of triaxial asteroids on highly eccentric
($e > 0.9$) orbits around stars, or, equivalently, highly eccentric triaxial satellites around planets. The method
relies on computing the velocity update after each pericenter passage -- not through direct numerical integration, but
rather with a surrogate function which accurately mimics the stochastic time evolution. This deterministic function
$d_\omega(\theta,\omega)$ is usually smooth (Fig. \ref{dom.fig}), enabling one to create an interpolation function 
for the integration (Fig. \ref{int.fig}).

We used the method to study the spin evolution of 2006 HY51 -- the asteroid with the highest known 
orbital eccentricity ($e=0.9684$) -- over 2 Gyr, and found that it does not spin itself apart. The maximum
spin rates seen in Gyr-long simulations are an order of magnitude lower than the estimated fission
threshold, which corresponds to $P_{\rm rot} \simeq 3$ hr \citep{ver+20}. Our
simulations did not include the possibility of chance encounters with Mercury, Venus, Earth or Mars during
the 2 Gyr evolution of 2006 HY51. Even if a close encounter did occur, any changes to the rotation speed 
of the asteroid would be small and quick, and those types of perturbations will not likely trigger a 
sequence of spin updates leading to breakup.

Our result suggests that (i) no other asteroids currently seen in the solar system are in danger of 
radiation-less rotational fission within the age of the solar system, and that (ii) the epoch when 
the asteroids achieved their current orbits
cannot be constrained by considering this type of breakup. However, primordial asteroids with 
eccentricities higher than that of 2006 HY51 could
have broken up and no longer be visible. In principle, for a given population and migration model of 
the early solar system \citep{nesvorny2018}, the population of minor planets perturbed onto highly eccentric
orbits can be time-evolved with our method to determine the fraction which break up, and the timescale
for doing so.

This study also confirms that the chaotic rotation of small-impact parameter asteroids in Table \ref{obj.tab}
is subject to abrupt changes at perihelion passages, which should be measurable. The median spin period
of the 2006 HY51, as follows from our long-term simulations, is 3 -- 4 d, but it may take any values
in a very wide range with periods as short as 1.5 d. If the period is close to this median value today,
the next perihelion passage may result in a jump of up to 0.7 d in spin period. It would be important
to observe this update to verify the prediction. The required determination of the spin period before
and after the nearest perihelion may be challenging, however, due to the faintness of these objects.
They become bright only for a few weeks or even days when they are close to the Sun, both geometrically
and in the sky projection, which precludes ground-based observations. For example, the 2012 US68
asteroid, which is about 22 mag at the time of writing this paper, will become as bright as 17.6 mag
on 2020-May-12, and will be brighter than 18 mag for some 25 days around this date, but it will
also be close to the lower conjunction with the Sun. Thus, the light curves
of these objects can be determined only when they are far away from the perihelion, and hence are faint.

For triaxial moons in our solar system, rotational breakup would not be a consideration: all known
moons have orbital eccentricities less than that of Nereid ($e \approx 0.75$). Nevertheless, our method
might be useful to track the time evolution of moons which feature chaotic spin evolution 
\citep[][]{tarnopolski2017a,tarnopolski2017b}. Further, the as-yet-undetermined population of exo-moons might 
provide additional candidates which are suitable for application with our tool.

We discovered the existence of stable spin-orbit resonances at high prograde rates of rotation in the
simplified 1D model with a single polar torque component. In multiple simulations with a high-fidelity
simulation technique, 2006 HY51 is captured with a probability of 0.5 into one of such resonances
within 6.5 Myr, at which point chaotic evolution of rotation ceases and a periodic, small-scale
circulation begins. This find does not obviate the main conclusion that 2006 HY51 and similar
asteroids in high eccentricity orbits cannot be destroyed by spin, because the resonances are
still an order of magnitude short of the break-up velocity. It is on our to do list to investigate if
the fascinating high-spin resonances are still possible in the more realistic 3D regime when the obliquity
is nonzero and all three principal axes torques are engaged. Furthermore, even a small perturbation
from interaction with one of the inner planets can possibly remove the asteroid from resonance,
for it to start on another chaotic journey for millions of years.

Finally, our method for fast evolution may have important implications for white dwarf planetary systems,
where a variety of dynamical scenarios can perturb asteroids onto orbits with $e > 0.99$ \citep{veras2016}
with observable consequences \citep{juryou2014,vanetal2015,farihi2016,doyetal2019,vanetal2019,manetal2020}.  
The maximum pericenter at which rotational breakup could occur was estimated to be about 
0.015 au \citep{ver+20}, but that value was based on full integrations over just $10^2$ orbits. 
A comprehensive parameter space study with our fast integration method over $\sim 10^9$ orbits 
might reveal that this maximum pericenter is much greater.
   
Even if 0.015 au is taken as the tentative limiting value, one may compare it with the orbital pericenter of 
2006 HY51, which is 0.082 au. The gap between these two values is sizeable, and where the transitional 
value for destruction
lies in-between is unknown. However, sublimation effects from the Sun would extend further than those from typical white dwarfs, suggesting that this pericenter difference may not solely be due to fundamental properties of chaotic spin evolution. A search in the Horizons database for all asteroids with a closest perihelion distance $r_{\rm min}
<0.085$ but without a brightness limit reveals several more poorly known asteroids in the main belt
in somewhat closer orbits, with an apparent cutoff at $r_{\rm min} =0.07$ au. These objects have shorter orbital
periods than the 2006 HY51, with the closest orbit for the 2005 HC4 : $P=2.46$ yr, $r_{\rm min} =0.071$ au,
$a=1.823$ au, $e=0.9613$. If we also select comets with the same search criteria, 23 additional objects
emerge with perihelion separations much shorter than 0.07 au. Interestingly, there is an inverse correlation
between $r_{\rm min}$ and the orbital period in this sample, with the closest ``Sun-grazing" comets having 
periods of hundreds of years. For example, the comet Pereyra (C\/1963 R1) has $P=903$ yr and $r_{\rm min}=
0.005$ au, probably the closest perihelion in the Solar system. A much longer period implies a much slower
chaotic evolution of rotation velocity; still, our analysis indicates that such extreme comets may be transient and short-lived.

\section*{Acknowledgments}

DV gratefully acknowledges the support of the STFC via an Ernest Rutherford Fellowship (grant ST/P003850/1).

\end{document}